\documentclass[floatfix,twocolumn,aps,showpacs,showkeys,nofootinbib]{revtex4-1}

\usepackage{amssymb}
\usepackage{amsmath}
\usepackage{graphics}
\usepackage{epsfig}
\usepackage[dvips]{color}
\usepackage{bm}

\newcommand{\be}{\begin{equation}}
\newcommand{\ee}{\end{equation}}
\newcommand{\bea}{\begin{eqnarray}}
\newcommand{\eea}{\end{eqnarray}}
\newcommand{\beas}{\begin{eqnarray*}}
\newcommand{\eeas}{\end{eqnarray*}}

\begin{document}
\title{Magnetic catalysis of a charged Bose-Einstein condensate}
\author{Alejandro Ayala,$^1$ M. Loewe,$^{2,3}$ Juan Cristobal Rojas,$^4$
C. Villavicencio$^{5,6}$}
\affiliation{  $^1$Instituto de Ciencias
  Nucleares, Universidad Nacional Aut\'onoma de M\'exico, Apartado
  Postal 70-543, M\'exico Distrito Federal 04510,
  Mexico\\
  $^2$Facultad de F\1sica, Pontificia Universidad Cat\'olica de Chile,
  Casilla 306, Santiago 22, Chile\\
  $^3$Centre for Theoretical and Mathematical Physics, University of Cape  Town,
  Rondebosch 7700, South Africa\\
  $^4$ Departamento de F\'isica, Universidad Cat\'olica del Norte,
  Casilla 1280, Antofagasta, Chile\\
  $^5$
  Departamento de F\'isica and 
  Centro-Cient\'ifico-Tecnol\'ogico de Valpara\'iso,\\
  Universidad T\'ecnica Federico Santa Mar\'ia,
  Casilla 110-V, Valpara\'iso, Chile\\
  $^6$Universidad Diego Portales, Casilla 298-V, Santiago, Chile}

\begin{abstract}
We study the condensation phenomenon for a system of charged bosons in the
presence of an external magnetic field. We show that condensation happens for a
definite critical temperature instead of through a diffuse phase transition. The
essential ingredient, overlooked in previous analyses and accounted for in this
work, is the treatment of the plasma screening effects by means of resummation.
We compute the critical temperature, for the case in which the condensate
is made of charged pions and for typical densities found in compact
astrophysical objects, for small and large values of the magnetic field. 
We show that the magnetic field catalyzes the onset of condensation at very
small and at large values of the magnetic field, and that for intermediate
values the critical temperature for condensation is lower than for the zero
magnetic field case.
\end{abstract}



\maketitle

\section{Introduction}\label{I}

The possibility that a charged pion condensate may occur in the interior of
neutron stars has been repeatedly examined in the past. 
This possibility is raised by the large isospin imbalance between neutrons
and protons which favors reactions that make neutrons decay into negative
pions under appropriate conditions.
The equilibrium thermodynamic conditions obeyed by a pion condensed
state in dense neutron matter and in neutron stars have been
discussed long ago. In particular, Ref.~\cite{Baym} studies the
criteria for the appearance of pion condensation in neutron matter
in terms of the pion Green's function (for a general review on the
physics of neutron stars see Ref.~\cite{Glendenning}). 
The occurrence of a charged boson condensed phase without magnetic
fields has also been extensively discussed in the literature.
Refs.~\cite{Sonandothers, Kogut, Splittdorf, Herpay:2008uw}
study in-medium processes introducing an isospin chemical potential
$\mu _{I}$ at zero temperature in both phases ($ |\mu _{I} |
\gtrless m_{\pi }$, where $m_\pi$ is the pion mass), analyzing the
formation of a charged pion condensed phase.
 This phenomenon was discussed in electrically neutral dense
quark matter in Refs.~\cite{Ebert,Abuki,Andersen}.
 Finite temperature
corrections, in the frame of chiral perturbation theory, have been
considered in
Ref.~\cite{Villavicencio1}, extending the discussion also to
other condensates, like the chiral condensate or the axial-isospin
charge density condensate, in Ref.~\cite{Villavicencio2}.

The situation becomes even more interesting when considering that neutron
stars possess large magnetic fields whose effects should also
be included when studying the condensation conditions.
Recently, in Ref.~\cite{Son2} the role played by the coupling of $\pi^{0}$
to a magnetic field via the triangle anomaly has been considered, showing
the emergence of a  $\pi ^{0}$ domain wall for values of the magnetic field
strength $B$ larger than a certain critical value. This could also happen
for $(\eta, \eta ')$ states when $ B \sim 10^{7} - 10^{19}$ G.

Magnetic fields can also play an important role in the dynamics of systems
where charged pions are copiously produced, such as relativistic heavy-ion
(RHI) collisions.
Recently, the importance of large magnetic fields for the evolution
of QCD matter produced in noncentral RHI collisions has been
discussed in Ref.~\cite{kharzeev} as well as their influence on the
phase structure of QCD, with emphasis on the chiral symmetry
restoration and deconfining transitions. In Ref.~\cite{Mizher}, a
discussion of the effective potential in the framework of the linear
sigma model, coupled to quarks and/or Polyakov loop, suggests a
richer structure of the strong interactions like, for example, a
possible splitting between chiral symmetry restoration and
deconfinement in the presence of magnetic fields. The influence of
the external magnetic field on the formation of CP-odd domains in
RHI collisions has also been discussed in Ref.~\cite{Mizher2}.
A decrease in the confining critical temperature was found in
\cite{Fraga:2012fs}, where a hadron-quark transition was studied within the
MIT
bag model. 
 With the above ingredients put together, the
theoretical study of a charged boson condensate with a finite
chemical potential in the presence of magnetic fields becomes even
more relevant. 
Although this is an old problem, the results from
several approaches vary in their conclusions. For instance, it was
long ago argued that a nonrelativistic Bose-Einstein gas of charged
particles does not condense in the presence of a magnetic field,
regardless of how weak the field may be \cite{Schafroth}. 
This
result motivated the search for conditions where condensation could
take place with magnetic fields, in particular to study whether this
could happen extending the number of spatial dimensions \cite{May,
Daicic, Elmfors}. 
In the nonrelativistic case, treating the
dimensionality $d$ of the system as a continuous variable, it was
shown in Ref.~\cite{May} that condensation can happen only for
$d>4$. For pairs of bosons or fermions and in the relativistic case,
it was shown in Ref.~\cite{Daicic} that for the case when $d$ is
taken as an integer, condensation happens for odd $d\geq 5$. A
similar conclusion was reached in Ref.~\cite{Elmfors}, although
these authors also realized that the lowest Landau level can play
the role of the ground state to accommodate a large charge density
in the $d=3$ case.

The common feature of all of the above-mentioned analyses is the
definition of the condensation condition which is taken as the
equality of the chemical potential and the ground state energy.
However, in the presence of a magnetic field, this condition leads
to a divergence of the particle density for that state. Indeed,
since for a constant magnetic field the energy levels separate into
transverse and longitudinal (with respect to the magnetic field
direction) and the former are described in terms of discrete energy
levels, the divergence of the Bose-Einstein distribution when the
chemical potential is equal to the lowest energy level can only be
cured in a larger than $d=4$ number of spatial dimensions.

The implications of this condition were recognized in Ref.~\cite{Perez} where
it was argued that when the temperature $T$ is much lower than $eB$ one can already consider that
 the system occupies only the
lowest Landau level. This means that the value for the chemical potential to
compute the ground state density does not need to be equal to the lowest energy.
In this picture the occupation of this state occurs without the need of having a
critical temperature, that is, the system undergoes a {\it diffuse phase
transition}.

Nevertheless one can argue that if in the absence of a magnetic field the
system is already in the condensed phase with a macroscopic fraction of the
population occupying the lowest energy level, a slow turning on of the magnetic
field should not lead to the instantaneous destruction of the condensate. 
Put in
equivalent terms, the onset of condensation for small magnetic fields should be
a phenomenon that takes place at a given critical temperature $T_c$ since it
does so in the limit of a vanishing magnetic field and the presence of a small one
cannot drastically change the picture. 
To implement this idea, one should keep
in mind that the chemical potential is not a number that can arbitrarily be
set to
take a specific value but rather, a function of the thermodynamic variables such
as temperature and density. 
Its value should be determined by demanding that
the ground state is populated by a finite charge density. 
The missing ingredient
that bridges the gap in the analysis is to consider the plasma screening effects,
which are of course needed since we are dealing with infrared phenomena
where {\it the effective mass} is small or may even vanish.

In this work we study the conditions for the onset of a condensed
phase for a charged boson system, in the presence of an external
magnetic field. To mimic the situation where there is an isospin
imbalance,  we introduce a finite  chemical potential $\mu$. 
For the
description, we resort to model the boson system in terms of a
theory of a charged scalar with quartic self-interactions. 
We show
that for small and large values of the magnetic field, the system
presents the magnetic catalysis phenomenon~\cite{catalysis}; that
is, that the formation of the condensate is favored by the presence
of the magnetic field.
This phenomenon has also been found in the context of the
Nambu-Jona-Lasino model at $T=0$~\cite{Klevansky2} and in (2+1)
dimensions both at $T=0$ and $T\neq 0$~\cite{Klimenko}, where it was
shown that even the presence of an arbitrary small magnetic field
breaks the chiral invariance of the models.
A main result of our work is to show that when including the plasma
screening effects, there is a well-defined
critical temperature associated with the onset of condensation.
A similar calculation, using optimized perturbation theory, albeit without
the introduction of a chemical potential, was done in Ref.~\cite{duarte}.
The authors found that the phase transition is always second order and the
magnetic catalysis phenomenon is present for all values of the magnetic
field. They also found that the critical temperature increases with
increasing values of the magnetic field.

 The work is organized as follows: In Sec.~\ref{II} we find the
lowest energy state where condensation happens and define the order
parameter for the transition. In Sec.~\ref{III} we compute the
one-loop corrections to the grand potential and set up the
discussion for the onset of the condensation phenomenon in terms of
the existence of a large but finite charge density in the ground
state. In Sec.~\ref{IV} we revisit the description of the onset of
condensation when corrections from interactions are accounted for.
We take the limit $B\to 0$ and point out the need to include plasma
screening effects by means of resummation, even in this case.
In Sec~\ref{V} we explicitly compute the resummed
self-energy for finite $B$ in the low temperature approximation in
the limits of small and large magnetic fields. This self-energy is
then used in Sec.~\ref{VI} to compute the critical temperature for
condensation when the charged bosons are taken as pions, for typical
densities in compact astrophysical objects such as neutron stars. We
finally summarize and conclude in Sec.~\ref{VII}.

\section{Order parameter}\label{II}

We want first to define the order parameter that describes the condensation
transition. 
This is a delicate task since (as we will show) in the presence of a
magnetic field, the condensate does not correspond to a spatially uniform
state.
Let us start by introducing the Lagrangian representing a charged scalar field
$\phi$ with finite chemical potential $\mu$ interacting with a uniform external
magnetic field $\bm{B}$ oriented in the $z$ direction. Working in the
symmetric gauge, the vector potential corresponding to the given magnetic field
can be written as
\bea
   {\bm{A}} =\frac{1}{2} {\bm{B}}\times{\bm{r}}.
\eea In Euclidean space and after introducing the magnetic field by
means of the minimal substitution we get \bea
   {\cal L}_\mathrm{E}&=&(\partial_\tau+\mu)\phi^*(\partial_\tau-\mu)\phi
   +|({\boldsymbol{\nabla}}-iq{\bm{A}})\phi|^2
   \nonumber\\
   &+&m^2|\phi|^2+\frac{\lambda}{4}|\phi|^4+\delta{\cal L},
\label{LE}
\eea
where $\delta{\cal L}$ contains the ultraviolet counterterms and $q$ is the
(positive) charge associated to the field $\phi$.  
We want to describe the
situation where for a given value of the chemical potential, the system
develops a superfluid phase characterized by a boson condensate  described
by a classical field $\phi_c$; namely, that the field can be expressed as
\bea
   \phi=\frac{1}{\sqrt{2}}\phi_c +\tilde\phi,
\label{field}
\eea
where $\tilde\phi$ is the quantum field, as referred from the real
classical ground state $\phi_c$. In the presence of an external magnetic
field, the classical equation of motion does not allow a constant value for
$\phi_c$~\cite{Harrington} and thus $\phi_c$ cannot simply be taken as the
order parameter for the transition. In order to see how one can proceed in
such a situation, let us find the ground state. Using the Lagrangian in
Eq.~(\ref{LE}) we obtain the classical action
\bea
   \Gamma_c &=&\beta\int
   d^3x\bigg\{
   \frac{1}{2}\phi_c(-\bm{\nabla}^2+q^2\bm{A}^2+m^2-\mu^2)\phi_c
   \nonumber\\
   &+&
   \frac{\lambda}{16}\phi_c^4\bigg\}
\label{Gamma_c}
\eea
where we have discarded a surface term after integration by parts  and
$\beta=1/T$. When the space boundary is not strictly taken at infinity, the
surface term does not vanish; however, given the form of the classical
solution [see Eq.~(\ref{ground}) below] the boundary contribution can be
neglected for a sufficiently large volume.
We first look for a solution for the free case $(\lambda =0)$; thus, the
eigenvalue problem
 for the classical equation of motion becomes
\bea
   \left[-\nabla^2+(qB)^2(x^2+y^2)/4+m^2-\mu^2\right]\phi_c = {\cal E}^2\phi_c
\label{class}
\eea
which is recognized as a two-dimensional harmonic oscillator whose
eigenvalues are given by
\bea
   {\cal E}^2_{l}(p_z) = p_z^2 +m^2+(2l+1)qB - \mu^2,
\label{eigenvalues}
\eea
where $l\geq 0$ labels the Landau level. Let us specialize to the lowest
energy state. This corresponds to $l=0$ and $p_z =0$ for which the solution
can be written as~\cite{Brito}
\bea
   \phi_c = v_0e^{-qB(x^2+y^2)/4},
\label{ground}
\eea
where $v_0$ can be determined from the normalization condition. 
The corresponding ground state energy, or effective mass squared, is given
by
\bea
   {\cal E}^2_{0}(0) = m^2 + qB - \mu^2.
\label{effmass}
\eea
As anticipated, the solution in Eq.~(\ref{ground}) is not spatially uniform. In order to define an appropriate
order parameter, we first normalize the solution over a given spatial volume $V$. This procedure involves finding
 the average over $V$ of $\phi_c^2$ defined as
\bea
   \langle \phi_c^2\rangle = \frac{1}{V}\int d^3x \ \phi_c^2.
\label{average}
\eea
From Eqs.~(\ref{ground}) and~(\ref{average}) one finds
\bea
   \langle \phi_c^2\rangle =v_0^2\left( \frac{1-e^{-\Phi/2\Phi_0}}{\Phi/2\Phi_0}\right),
\label{averageexpl}
\eea
where $\Phi\equiv BA$ is the magnetic flux passing through the transverse area
$A$ and $\Phi_0\equiv\pi/q$ is the quantum magnetic flux. Notice that for the
analysis, neither $A$ nor $B$ can be taken as changing independently but instead
that $\Phi$ should be considered as the relevant variable.
Also, notice that the term between the parentheses in
Eq.~(\ref{averageexpl}) goes
to 1 as $\Phi\rightarrow 0$, as expected.
The requirement to obtain the effective mass squared
independent of the magnetic flux, leads us to consider
$\bar{\phi}_c\equiv\sqrt{\langle \phi_c^2\rangle}$ as the order parameter to
describe the condensation transition. This is determined as follows.
% Also, notice that the term between parenthesis in Eq.~(\ref{averageexpl})
% vanishes as $\Phi\rightarrow \infty$, however, it goes to 1 as $\Phi\rightarrow
% 0$. This property leads us to consider $\bar{\phi}_c\equiv\sqrt{\langle
% \phi_c^2\rangle}$ as the order parameter to describe the condensation
% transition.
In terms of $\bar{\phi}_c$ the ground state solution reads as
\bea
   \phi_c = \bar{\phi}_c \left( \frac{\Phi/2\Phi_0}{1-e^{-\Phi/2\Phi_0}}\right)^{1/2}e^{-qB(x^2+y^2)/4}.
\label{groundexpl}
\eea
We now look for the value of $\bar{\phi}_c$ that minimizes the classical action, this time accounting
 for the effects of the self-interaction ($\lambda\neq 0$). Substituting Eq.~(\ref{groundexpl}) into
  Eq.~(\ref{Gamma_c})  we get
\bea
   \Gamma_c=\beta V \left[
   \frac{1}{2}(qB +m^2-\mu^2)\langle\phi_c^2\rangle
   +\frac{\lambda}{16}\langle\phi_c^4\rangle
\right],
\label{Gamma-vac}
\eea
where in general one defines
\bea
   \langle\phi_c^{n}\rangle&\equiv&  \frac{1}{V}\int d^3x \ \phi_c^n\nonumber\\
   &=&v_0^n\left(\frac{1-e^{-n\Phi /4\Phi_0}}{n\Phi /4\Phi_0}\right).
\eea
Therefore, the nontrivial minimum of Eq.~(\ref{Gamma-vac}) is found for a
value of the
 order parameter $\bar\phi_c$ given explicitly by
\bea
   \bar{\phi}_{c0}
   &=&4\sqrt{\frac{\mu^2 - m^2 -qB}{\lambda}}
   \left(\frac{\Phi_0}{\Phi}\right)^{1/2}\nonumber\\
   &\times&\left(\frac{1-e^{-\Phi /2\Phi_0}}{1+e^{-\Phi /2\Phi_0}}\right)^{1/2}
   \theta (\mu^2 - m^2 -qB).
\label{min-class}
\eea
Notice that for a given value of $B$ (and of $\Phi$), since the thermal
occupation of the ground state grows when $\mu^2 \rightarrow m^2 + qB$,
then Eq.~(\ref{min-class}) means that a macroscopic fraction of the charged
particles will occupy the condensed state as $\bar\phi_{c0}\rightarrow
0^+$. 
One can expect that the same is true when considering a higher order
$n$ in the perturbative expansion of the effective action; that is, that
the superfluid transition is signaled by the condition that
$\bar\phi_{cn}\rightarrow 0^+$. 
We now proceed to find how this condition
is realized at order $n=1$.

\section{One-loop effective potential}\label{III}

The corrections to the value that minimizes the action are obtained from the grand potential.
For the theory at hand, this is given by
\bea
   \Omega(\bar \phi_c) = -\frac{1}{\beta V} \ln \int D\tilde\phi^*
   D\tilde\phi~e^{-S[\phi^*,\phi]},
\label{grandpot}
\eea
where $S[\phi^*,\phi]$ is the action defined in Eq.~(\ref{LE}) and $\phi$ and $\tilde{\phi}$ are
 related as in Eq.~(\ref{field}).
At one-loop order, the grand potential has the explicit expression
\begin{equation}
   \Omega(\bar \phi_c) = (m_B^2-\mu^2)\langle \phi_c^2\rangle
   +\frac{\lambda}{4}\langle \phi_c^4\rangle
   +\frac{1}{2 \beta V}\ln\det\mathbb{D}^{-1},
\label{grandoneloop}
\end{equation}
where the inverse propagator matrix operator is defined as
\bea
 \!\!\! \!\!\! \mathbb{D}^{-1}=\left(
                                      \begin{array}{cc}
                                     -{\cal D}^2_- +m^2+
   \frac{\lambda}{2}\phi_c^2 & \frac{\lambda}{4}\phi_c^2\\
                                     \frac{\lambda}{4}\phi_c^2 & -{\cal
   D}^2_++m^2+\frac{\lambda}{2}\phi_c^2
                                     \end{array}
                                \right),
\label{matrixprop}
\eea
with
\bea
   {\cal D}_\pm^2 =(\partial_\tau \pm\mu)^2+(\bm{\nabla}\pm iq\bm{A})^2.
   \label{Dpm}
\eea
Hereafter we use the notation
\bea
   m_B\equiv \sqrt{m^2+qB},
\label{mB} \eea for the effective mass in the lowest Landau level.

The fact that the operators ${\cal D}_\pm^2$ depend on the coordinates makes it difficult to find the
functional determinant. Nevertheless, since our interest is to explore the condensation phenomenon near the
phase transition, we can resort to expand the grand potential in powers of the order
parameter near the value $\bar{\phi}_{c1}$ that minimizes it:
\bea
   \Omega(\bar \phi_c)\approx \Omega (\bar \phi_{c1}) +\frac{1}{2}(\bar \phi_c - \bar \phi_{c1})^2
   \left(\frac{\partial^2\Omega}{\partial \bar \phi_c^2 }\right)
   \Big|_{\bar \phi_c = \bar \phi_{c1}}
\label{Omegapporx}
\eea
The factor
\bea
   {\cal{M}}^2(\bar\phi_{c1})\equiv \left(\frac{\partial^2\Omega}{\partial \bar
\phi_c^2}\right)\Big|_{\bar \phi_c = \bar \phi_{c1}}
\label{effmassgrandpot}
\eea
can be regarded as an effective mass squared for the quantum field
$\tilde{\phi}$. Following the discussion at the end of Sec.~\ref{II}, let us
explore the behavior of $\Omega (\bar \phi_{c1})$ and  ${\cal{M}}^2(\bar
\phi_{c1})$ near the superfluid phase transition where we expect that $\bar
\phi_{c1}\rightarrow 0^+$. Notice also that ${\cal{M}}^2$ represents the
curvature of the grand potential in the direction of $\bar \phi_c$. For a given
value of $B$ (and of $\Phi$) the system is in the normal phase when the
curvature is positive. As the occupation number of the ground state increases,
the curvature should tend to change sign. Therefore the transition to the
superfluid phase is also signaled by the condition ${\cal{M}}^2\rightarrow 0^+$.
The explicit one-loop expressions for $\Omega (0)$ and ${\cal{M}}^2(0)$ are
\bea
   \Omega (0) &=& T\sum_{n=-\infty}^\infty\int\frac{d^3p}{(2\pi)^3}\ln D^{-1}\nonumber\\
   {\cal{M}}^2(0)&=&m_B^2-\mu^2 + \lambda T\sum_{n=-\infty}^\infty\int\frac{d^3p}{(2\pi)^3} D,
\label{MandOexpl}
\eea
where $D$ is the propagator for a charged scalar in the presence of a constant magnetic field.
We use the expression for $D$ obtained in the Schwinger proper time method, given by
\bea
   D=\int_0^\infty \frac{ds}{\cosh (qBs)}
   e^{-s\left[(\omega_n-i\mu)^2+p_z^2+m^2+p_\perp^2\frac{\tanh
   (qBs)}{qBs}\right]},\nonumber\\
\label{schwprop}
\eea
where $p_\perp^2$ represents the square of the components of ${\bm{p}}$
transverse to the direction of the magnetic field and $\omega_n$ is a boson
Matsubara frequency. Also, in writing Eq.~(\ref{schwprop}) we have ignored a
phase factor which does not contribute when considering closed loop expressions.
Note that the one-loop correction in ${\cal{M}}^2(0)$ is in fact the
self-energy. This is not the case when one considers higher loop corrections.
Carrying out the integrations over the transverse components we get
\bea
   \Omega(0) &=& \frac{qBT}{2\pi}
   \sum_{l,n}\int\frac{dp_z}{2\pi}\ln(P^2+m^2)\nonumber\\
   {\cal{M}}^2(0) &=& m_B^2-\mu^2
   + \lambda\frac{qBT}{2\pi}\sum_{l,n}\int\frac{dp_z}{2\pi}
   \frac{1}{P^2+m^2}\qquad
\label{afterptint}
\eea
where
\bea
   P^2 = (\omega_n-i\mu)^2+p_z^2+qB(2l+1)
\label{capitalP}
\eea
with $l \geq 0$ being the index labeling the Landau levels.

Let us now separate the $T=0$ (purely magnetic field) contribution from
the thermal dependence by writing
\bea
   \Omega(0) &=& \Omega_B + \Omega_{T,B}\label{Omega_0-1}\nonumber\\
   {\cal{M}}^2(0) &=& m_B^2-\mu^2+\Pi_B+\Pi_{T,B}
\label{Omega_2-1}
\eea
where the subscripts $T$ and $B$ denote the temperature and
magnetic field-dependent contributions.

Let us first compute the purely magnetic field contributions. As shown in the Appendix,
these contributions are easily obtained using dimensional regularization. 
They can be expressed in terms of the Hurwitz zeta function $\zeta(s,u)$
and are given explicitly by
\bea
   \Omega_B &=& \frac{m^4}{(4\pi)^2}\Bigg[
   \frac{1}{2}\ln\left(\frac{2qB}{m^2}\right)\nonumber\\
   &+&
   \left(\frac{2qB}{m^2}\right)^2
   \zeta'\left(-1, \frac{1}{2}+\frac{m^2}{2qB}\right)\Bigg]\nonumber\\
   \Pi_B &=&
   \frac{\lambda m^2}{(4\pi)^2}\Bigg[
   1+\ln\left(\frac{2qB}{m^2}\right)
   \nonumber\\
   &+&
   \left(\frac{2qB}{m^2}\right)
   \ln\left\{
   \Gamma\left(\frac{1}{2}+\frac{m^2}{2qB}\right)/\sqrt{2\pi}
   \right\}
   \Bigg],
\label{OmegaPiexpl}
\end{eqnarray}
where $\zeta'(s,u) = \partial\zeta(s,u) /\partial s$ and $\Gamma (u)$ is the
gamma function. In writing Eq.~(\ref{OmegaPiexpl}) we have chosen
the renormalization scale in the  ${\overline{\text{MS}}}$ scheme as
$\Lambda_{\overline{\text{MS}}}=me^{-1/2}$. With this choice, one gets a
vanishing contribution in the limit $B\to 0$.

As is also shown in the Appendix, the thermal contribution to $\Omega(0)$ and
${\cal{M}}^2(0)$ can be expressed as
\bea
   \Omega_{T,B} &=&
   -\frac{qB}{4\pi^2}\sum_{n=1}^\infty \cosh(\beta\mu n)
   \int_0^\infty \frac{ds}{s^2}
   \frac{e^{-s m_B^2- \beta^2 n^2 / 4s}}{1-e^{-2qB s}},\nonumber\\
   \Pi_{T,B} &=&
   \lambda\frac{qB}{4\pi^2}\sum_{n=1}^\infty \cosh(\beta\mu n)
   \int_0^\infty \frac{ds}{s}
   \frac{e^{-s m_B^2- \beta^2 n^2 / 4s}}{1-e^{-2qB s}}.
   \nonumber\\
\label{Pi_TB}
\eea
Since condensation is a low temperature phenomenon, let us
approximate Eq.~(\ref{Pi_TB}) in the limit where $T\ll m_B$.  In this case,
the
integrals can be computed using the steepest descent method and the result can be expressed as
\bea
   \Omega_{T,B} &\approx& -2\pi m_B^4 \tau^{5/2}
   \left[ \gamma Li_{3/2}(z)+\sum_{n=1}^\infty\frac{z^n}{n^{5/2}}
   \frac{n\gamma}{e^{n\gamma}-1}
   \right]
   \nonumber\\
   &+&
   \{ \mu\to-\mu\},\nonumber\\
   \Pi_{T,B} &\approx&
   \frac{\lambda}{2} m_B^2 \tau^{3/2}
   \left[ \gamma Li_{1/2}(z)+\sum_{n=1}^\infty\frac{z^n}{n^{3/2}}
   \frac{n\gamma}{e^{n\gamma}-1}
   \right]
   \nonumber\\
   &+&
   \{ \mu\to-\mu\},
   \label{Pi_TB1}
\eea
where the polylogarithm function is defined as
\bea
   Li_{s}(z)\equiv\sum_{n=1}^\infty\frac{z^n}{n^s}
\label{polylog}
\eea
and the fugacity $z$, scaled temperature $\tau$ and scaled magnetic field $\gamma$ are defined as
\bea
   z &\equiv& e^{(\mu-m_B)/T},\\
   \tau &\equiv& \frac{T}{2\pi m_B},\\
   \gamma &\equiv&\frac{qB}{m_B T},
\label{scaled}
\end{eqnarray}
respectively.
Notice that the sum over the index $n$ in Eq.~(\ref{Pi_TB1}) corresponds to
a sum over Matsubara frequencies.
Also, in writing Eq.~(\ref{Pi_TB1}) from Eq.~(\ref{Pi_TB}), we have
explicitly separated the contribution from the lowest Landau level---whose
expression is given in terms of the polylogarithm function---from the
contribution of the rest of the energy levels. This separation proves
useful since the contribution from the lowest Landau level, unlike that
from the rest of the levels, is strongly infrared divergent near the phase
transition and must be treated separately~\cite{Perez}.

Recall that the charge density is defined as
\bea
   \rho=-\frac{\partial \Omega (\bar\phi_c)}{\partial\mu}.
\label{chargedendef}
\eea
Near the superfluid transition where ${\cal{M}}^2(0)\to 0$ one can compute the charge density by
considering only the term $\Omega (0)$ in the grand potential, therefore, at one-loop order, the
thermal part of the charge density is obtained from the first line of
Eq.~(\ref{Pi_TB1}) as
\bea
   \rho &\approx&
   m_B^3 \tau^{3/2}
   \left[ \gamma Li_{1/2}(z)+\sum_{n=1}^\infty\frac{z^n}{n^{3/2}}
   \frac{n\gamma}{e^{n\gamma}-1}
   \right]
   \nonumber\\
   &-&
   \{ \mu\to-\mu\}.
\label{rho1}
\eea
Since the phase transition happens when $\mu\sim m_B$
the terms with $\{ \mu\to-\mu\}$ in Eqs.~(\ref{Pi_TB1}) and~(\ref{rho1}) are negligible, given that they become
proportional to powers of the factor
$\exp(-(m_B+\mu)/T)\sim\exp(-2m_B/T)\ll 1$,
and thus hereafter we ignore them. Notice that with this approximation, the
second equation of Eq.~(\ref{Pi_TB1}) and Eq.~(\ref{rho1}) imply that 
$\Pi_{T,B}$ and
$\rho$ are proportional.
We have now set up the stage to discuss
the condensation phenomenon in terms of a finite charge density in the ground
state.

\section{Bose-Einstein condensation revisited}\label{IV}

The existence of a critical temperature and a critical chemical potential
indicates that the system of charged bosons reaches a kind of saturation
where the occupation of the ground state becomes important.
 This saturation leads to the superfluidity phenomenon. From the
computational point of view, the condensation conditions are searched for
from the values of the parameters that minimize
the vacuum energy. Given that the description of the onset of condensation may
be obscured by the existence of infrared divergent quantities and in order to
gain insight, let us first revisit how these conditions are found at tree
and
one-loop level.

At tree level, the condensation condition is given by $\mu=m_B$. This condition implies
 the vanishing of the mass term ${\cal M}^2(0)$ in the grand potential, as
can be seen from Eq. (\ref{Omega_2-1}). If we now consider the one-loop correction, from
 the second equation of Eq.~(\ref{Pi_TB1}) and from the limiting behavior
of $Li_{1/2}(z)$ for
  $\mu\to m_B$ $(z\sim 1)$ and $\mu\ll m_B$ $(z\ll 1)$
\bea
   Li_{1/2}(z)\approx\left\{
   \begin{array}{cl}
   \sqrt{\frac{\pi}{1-z}}+\zeta(\frac{1}{2}) & \text{for}~~ z\lesssim 1\\
   z & \text{for}~~ z\ll 1,
   \end{array}
   \right.
\label{limit}
\eea
we see that ${\cal M}^2(0)$ diverges when $\mu\to m_B$.
This behavior has been interpreted as the impossibility of the existence of
a superfluid state in the presence of an external magnetic
field~\cite{Schafroth}. However, this
usual prescription for the onset of condensation, is not adequate in the
presence of an external magnetic field.

Consider the situation in the absence of a magnetic field. Intuitively, once the
superfluid phase is established, it is difficult to imagine that this can be
instantaneously destroyed by the turning on of an arbitrary small external
magnetic field. One would expect that in case the magnetic field destroys the
superfluid state, when the field is small, the condensed state should be
restored for a different
temperature. Similar considerations were made for the case of a
noninteracting gas in
Ref.~\cite{Perez} albeit for a high external magnetic field, in the limit $T\to
0$. In such case it was shown that the charge density in the normal phase
vanishes, i.e., all the charges populate the superfluid phase. However, since
the physical conditions require to have a finite charge density, the chemical
potential---which depends on temperature as well as on this charge
density---does not reach the value that corresponds to the ground state
energy and
therefore no infrared divergence occurred. It is important to emphasize that
Ref.~\cite{Perez} suggests that under such conditions, there is no definite
critical temperature associated with the superfluid transition.

To continue gaining insight, let us keep on analyzing the case with zero
external magnetic field. From Eq.~(\ref{rho1}), the charge density for
$\gamma\to 0$
%in the normal phase for $\mu\sim m$
becomes
\bea
   \rho \approx
   \left(\frac{mT}{2\pi}\right)^{3/2}
   Li_{3/2}(e^{(\mu-m)/T}).
\label{rho2}
\eea
The function $Li_{3/2}(z)$ is well defined for $z\leq 1$, although its
derivative diverges at $z=1$. Notice that there is
no analytical continuation for $Li_{3/2}(z)$ which gives a real result for $z>1$.
We have explicitly
\bea
   Li_{3/2}(z)\approx\left\{
   \begin{array}{cl}
   \zeta(\frac{3}{2})+\sqrt{4\pi(1-z)} & \text{for}~~ z\lesssim 1\\
   z & \text{for}~~ z\ll 1.
   \end{array}
   \right.
\label{poly3/2expl}
\eea
Thus, the maximum value allowed for the chemical potential is $\mu_0=m$ $(z=1)$.
This result is interpreted as the saturation of the boson system that gives rise
to condensation at this critical value of the chemical potential.
Since the charge density is a conserved quantity, there must be a critical
temperature $T_0$ for which condensation takes place. In other words
$\mu_0=\mu(T_0,\rho)$. The condition to obtain the critical temperature comes from Eq.~(\ref{rho2})
 by setting $\mu_0=m$, resulting in
\bea
   T_0=\frac{2\pi}{ m}\left(\frac{\rho}{\zeta(\frac{3}{2})}\right)^{2/3}.
\label{T0}
\eea
This is the well-known result for the critical temperature of a
noninteracting boson
gas. For temperatures lower than the critical temperature, $T<T_0$, the gas can be
separated into two phases: the normal ($N$) phase and the superfluid ($S$)
phase.
The charge density splits into these two states $\rho=\rho_N+\rho_S$, where
the charge density in the normal phase $\rho_N$ is defined as the charge
density $\rho$, evaluated at the critical chemical potential $\mu_0$. In
the absence of the magnetic field this is given by
\bea
   \rho_N &=& \left(\frac{mT}{2\pi}\right)^{3/2}\zeta(3/2).
\label{rhoNorm}
\eea
The superfluid charge density corresponds to the difference between the total
charge density and the charge density in the normal phase, $\rho_S=\rho-\rho_N$,
for a fixed total charge density $\rho$.

When interactions are accounted for, the situation changes. The condition for
the phase transition is once again looked for from the vanishing of the
effective mass squared ${\cal {M}}^2(0)$. From Eq.~(\ref{Omega_2-1}) and the
second equation of Eq.~(\ref{Pi_TB1}), in the limit
$B\to 0$, the effective mass squared is given by
\bea
   {\cal M}^2(0) &=&
   m^2-\mu^2\nonumber\\
   &+&\frac{\lambda m^2}{2}\left(\frac{T}{2\pi m}\right)^{3/2}
   Li_{3/2}(e^{(\mu-m)/T})\nonumber\\
   &\stackrel{\mu\to m}{\longrightarrow}& \frac{\lambda
m^2}{2}\left(\frac{T}{2\pi m}\right)^{3/2}\zeta(3/2),
\label{effmassB0}
\eea
which shows that ${\cal M}^2(0)$ cannot vanish for $\mu\leq m$.  Moreover, since
the function $Li_{3/2}(z)$ cannot be analytically continued to real values for
$z>1$, there is no physical solution that sets ${\cal M}^2(0)=0$ even if we were
to consider $\mu>m$.

The above results show the need of an extra ingredient already for $B=0$.
In this case, the solution is well known: since the physical conditions
require the effective mass squared to vanish, plasma screening effects need
to be accounted for by means of resummation. Examples of the importance to
include resummation effects have been recently discussed for systems
subject to the influence of an external magnetic field. For instance in
Ref.~\cite{Ayala1}  the temperature dependent effective potential for a
scalar theory, similar to the case here discussed, was considered. A
resummation of ring diagrams turns out to be extremely important to
understand the appearance of a first order phase transition. The scenario
has also been considered in other theories such as the linear sigma
model~\cite{Ayala2} and the standard model, during the electroweak phase
transition~\cite{Ayala3}. In all these cases, a second order phase
transition turns into a first order one.

In the present context inclusion of resummation effects means that the self-energy should be computed
 self-consistently as
\bea
   \Pi\to
   \bar\Pi=\lambda T\sum_{n=-\infty}^\infty\int\frac{d^3p}{(2\pi)^3}
   D(m^2\rightarrow m^2 +
   \bar\Pi).
\label{piselfconst}
\eea
To simplify the calculation, and in case the coupling $\lambda$ is not too
large, it is customary to substitute inside the argument of the propagator in
Eq.~(\ref{piselfconst}), $\bar\Pi\rightarrow\Pi_1$. For the low temperature
expansion for $B=0$ and
when $\mu=m$, we see from the second equation of Eq.~(\ref{Pi_TB1}) and
Eq.~(\ref{poly3/2expl}) that in this case
\bea
   \Pi_1
   = \frac{\lambda m^2}{2}\left(\frac{T}{2\pi m}\right)^{3/2}\zeta(3/2).
\label{pi1}
\eea
When the coupling $\lambda$ is not small, a full self-consistent treatment is
needed.

In the following section we explain in detail the resummation procedure
for a finite external magnetic field. From there, the case treated in this
section is obtained as the limit with $B\to 0$.

\section{Resummation at finite $B$}\label{V}

As was mentioned in the previous section, in order to consistently
compute the critical temperature when interactions are accounted for, it
 is necessary to consider resummation. Inclusion of these effects allows
  us to find the chemical potential, beyond its tree level value $\mu=m$, by
   consistently accounting for plasma screening.
   The need for resummation is even more dramatic for $B\neq 0$, since
when
    ignoring screening and $\mu\rightarrow m_B$ the charge density and the
     thermal contribution to the self-energy both diverge in the
     infrared. In Ref.~\cite{Fendley}  the thermal effective potential in the
$\lambda \phi ^{4}$ theory was computed by a resummation of the ring
diagrams.
Effectively, the resummation was done through a solution of the
renormalization group equation.
Resummation results in a shift of $m^2$ in the one-loop correction of the grand potential
\bea
   \Omega(0)\to \Omega_r(0) &=& \left.\Omega_B+\Omega_{T,B}\right|_{m^2\to
m^2+\bar\Pi}
   \nonumber\\
   {\cal M}^2(0)\to {\cal M}^2_r(0) &=& m_B^2-\mu^2
   \nonumber\\&&
   +\left.\Pi_B+\Pi_{T,B}\right|_{m^2\to
   m^2+\bar\Pi},
\label{potres}
\eea
where the subscript $r$ is to emphasize that resummation effects are included.
As a result, the charge density becomes
\bea
  \rho = -\frac{\partial\Omega_r(0)}{\partial\mu}
  \label{chargeres}
\eea
The resummed self-energy $\bar\Pi$ is obtained self-consistently
\bea
   \bar\Pi &=& \left.\lambda\frac{qBT}{2\pi}\sum_{l,n}\int\frac{dp_z}{2\pi}
   \frac{1}{P^2+m^2+\bar\Pi}\right|_{\mu=m_B}
   \nonumber\\&-& \{T\to 0\},
\label{selfressumed}
\eea
where $P^2$ is defined in Eq. (\ref{capitalP}).
We consider only the
thermal contribution, as indicated by subtracting the $\{T\to 0\}$ term. Also
since we are interested in computing this self-energy near the phase transition,
we consider its value for $\mu=m_B$. The self-consistent equation is reduced to
finding the solution to the expression
\bea
   \bar\Pi =\left.\Pi_{T,B}\right|_{m^2\to m^2+\bar\Pi,~\mu=m_B}.
\label{barPi-2}
\eea
For the rest of this section, we concentrate on finding an explicit expression
for
$\bar\Pi$ in some limits.

In order to simplify the self-consistent equation, recall that
since we want to include the effect of thermal fluctuations coming from the
resummation of the so-called {\it ring diagrams} for small temperatures, we
use
the low temperature approximation for $\Pi_{T,B}$ given in Eq.~(\ref{Pi_TB1}).
We can consider $\bar\Pi\ll m^2$ up to the leading order in the right-hand side
of Eq.~(\ref{barPi-2}). At leading order, we can set $\bar\Pi = 0$ in all the
terms, except in the argument
\bea
   z=e^{\left(\mu-\sqrt{m^2+qB}\right)/T}\Big|_{m^2\to m^2+\bar\Pi,~\mu\to\ m_B}
\label{zren}
\eea
of the polylogarithm function $L_{1/2}(z)$, since this diverges when
$\mu=m_B$ in the limit $\bar\Pi\to 0$. Therefore we set
\bea
   z &\to&
   e^{\left(m_B-\sqrt{m^2+\bar\Pi+qB}\right)/T}
   \nonumber\\
   &\approx& e^{-\bar\Pi/2m_BT}.
\eea
We explore the situation where $\bar\Pi\ll m_B T$, which
is a good enough condition to control the infrared divergences.
Then we can expand the fugacity as
\bea
   z\approx
   1-\Pi/2m_BT.
\label{zapprox}
\eea
Using the expression for $Li_{1/2}$ in Eq.~(\ref{limit}) we obtain a
simplified expression for the self-consistent equation
\bea
   \!\!\!\!\!\!\!\!\! \bar\Pi&\approx&
   \frac{\lambda }{2}m_B^2 \tau^{3/2}
   \left[ \gamma \sqrt{\frac{2\pi m_B T}{\bar \Pi}}
   +\sum_{n=1}^\infty\frac{1}{n^{3/2}}
   \frac{n\gamma}{e^{n\gamma}-1}
   \right].
   \label{barPi-3}
\eea
This is a third order equation in the variable $\sqrt{\bar\Pi}$. It can be solved
explicitly, though the solution
is given in terms of complicated
expressions. It is therefore more instructive to find the
solution for different values of
$T$ and $B$, in particular, the next-to-leading order correction in
$\lambda$ for the case $qB\ll m_BT$ here treated.
For this purpose, we expand Eq.~(\ref{barPi-3}) for $\gamma\rightarrow 0$.
Also, since we are close to $B=0$ we can write $\Pi_1$, given in Eq.~(\ref{pi1}), instead of $\bar\Pi$ in the
right-hand side of Eq.~(\ref{barPi-3}), resulting in
\begin{equation}
   \bar\Pi \approx
\frac{\lambda^{1/2} qB}{\sqrt{2\zeta(3/2)}}\left(\frac{T}{2\pi m}\right)^{1/4}
+\frac{\lambda m^2}{2}\left(\frac{T}{2\pi m}\right)^{3/2}\zeta(3/2).
\label{barPi_lowB}
\end{equation}
It can be shown that the first term on the right-hand side of Eq.~(\ref{barPi_lowB}) is the
 leading correction coming from
the lowest Landau level. Notice that this term also makes explicit the breaking of the perturbative
 regime as it is proportional to $\lambda^{1/2}$.

We emphasize that resummation is needed only for the case when considering
small values of the external magnetic field, compared with temperature.
In fact, notice that the nonthermal contribution to the self-energy $\Pi_B$
in
Eq.~(\ref{OmegaPiexpl}) is negative. Its limiting values can be approximated as
\bea
   \Pi_B\approx\left\{
   \begin{array}{cl}
   -\frac{\lambda}{6}\left(\frac{qB}{4\pi m}\right)^2
   & \text{for}~ qB<m^2\\
   -\frac{\lambda qB}{(4\pi)^2}\ln 2
   &\text{for}~ qB\gg m^2.
   \end{array}
\right.
\label{limitB}
\eea
Therefore, in this case there is always a sufficiently small temperature such
that $\Pi_B +\Pi_{T,B}<0$ for $\mu\lesssim m_B$ and the effective mass squared
${\cal M}(0)^2$ in Eq. (\ref{Omega_2-1}) can thus vanish.

\section{Critical temperature}\label{VI}

Having solved the infrared divergence problem by the introduction of
the resummation procedure in the self-energy, we now proceed to find
the critical temperature for the superfluid phase transition and 
explore how this critical temperature is modified as a function of
the strength of the magnetic field, compared to the zero field case.
As we pointed out in Sec.~\ref{IV}, the charge density is a
conserved quantity and therefore it cannot diverge. From the
resummed version of Eq.~(\ref{rho1}), that is, with the mass shifted
as $m^2\to m^2+\bar\Pi$, we can express the chemical potential in
terms of the charge density and the other parameters, namely,
$\mu=\mu(\rho,T,B)$. For a fixed charge density and a given value of
the magnetic field, the critical temperature $T_c(\rho,B)$ can be
found from the condition that sets the grand potential up to its
minimal value. Recall that for $B=0$ it is well known that the phase
transition is of the second order.
Even more, in the case of zero chemical potential, the
transition is always of second order \cite{duarte}.
Here we will assume that, for
$B\neq 0$ the phase transition continues being of the second order,
which is the expected behavior if we slowly turn on the magnetic
field.

\bea
   \left.{\cal M}_r^2(0)\right|_{T=T_c, ~\mu=\mu_c} &=& 0\nonumber\\
   \left.-\frac{\partial\Omega_r(0)}{\partial\mu}\right|_{T=T_c, ~\mu=\mu_c}
   &=& \rho,
\label{condcritTmu}
\eea
where $\mu_c(\rho,B)\equiv \mu(T_c,\rho,B)$.
We proceed to numerically solve Eq.~(\ref{condcritTmu}).

In the low temperature approximation that we are considering, the thermal
contribution to the self-energy is proportional to the charge density,
as can be seen from Eqs.~(\ref{Pi_TB1}) and (\ref{rho1}), upon neglecting the terms
with $\{\mu\to -\mu\}$. Therefore we can write
\bea
   \Pi_{T,B} = \frac{\lambda \rho}{2 m_B}.
\eea
The same is true after including resummation effects.
Since $\rho$ is constant, we can use the above result to immediately solve
for the critical chemical potential from the first line of
Eq.~(\ref{condcritTmu}), resulting in
\bea
   \mu_c \approx \left[
   m_B^2 +\Pi_B + \frac{\lambda\rho}{2 \sqrt{m_B^2+\bar\Pi}}
   \right]^{1/2}.
\label{muc}
\eea
Notice that now, finding the critical temperature becomes equivalent to
solving Eq.~(\ref{rho1}) with the replacement $m^2\to m^2+\bar\Pi$, where
for $T$ and $\mu$ use is made of their critical values $T=T_c,~ \mu=\mu_c$,
with $\mu_c$ given explicitly in Eq.~(\ref{muc}). Let us now find $T_c$
numerically. For this purpose, we consider a charged pion condensate in the
core of compact stars. The mass is given by $m=m_\pi\approx 140$ MeV. The
typical surface temperature for a cold neutron star is $T\sim 100$ eV.
Since the critical temperature $T_c$ for small $B$ is expected to be close
to the value $T_{c0}$ at zero magnetic field, we can use Eq.~(\ref{T0}) to
obtain the charge density that is needed for the onset of condensation at
these temperatures.
%%%%%%%%%%%%%%%%%%%%%%%%%%%%%%%%%%
\begin{figure}
   \includegraphics[scale=.64]{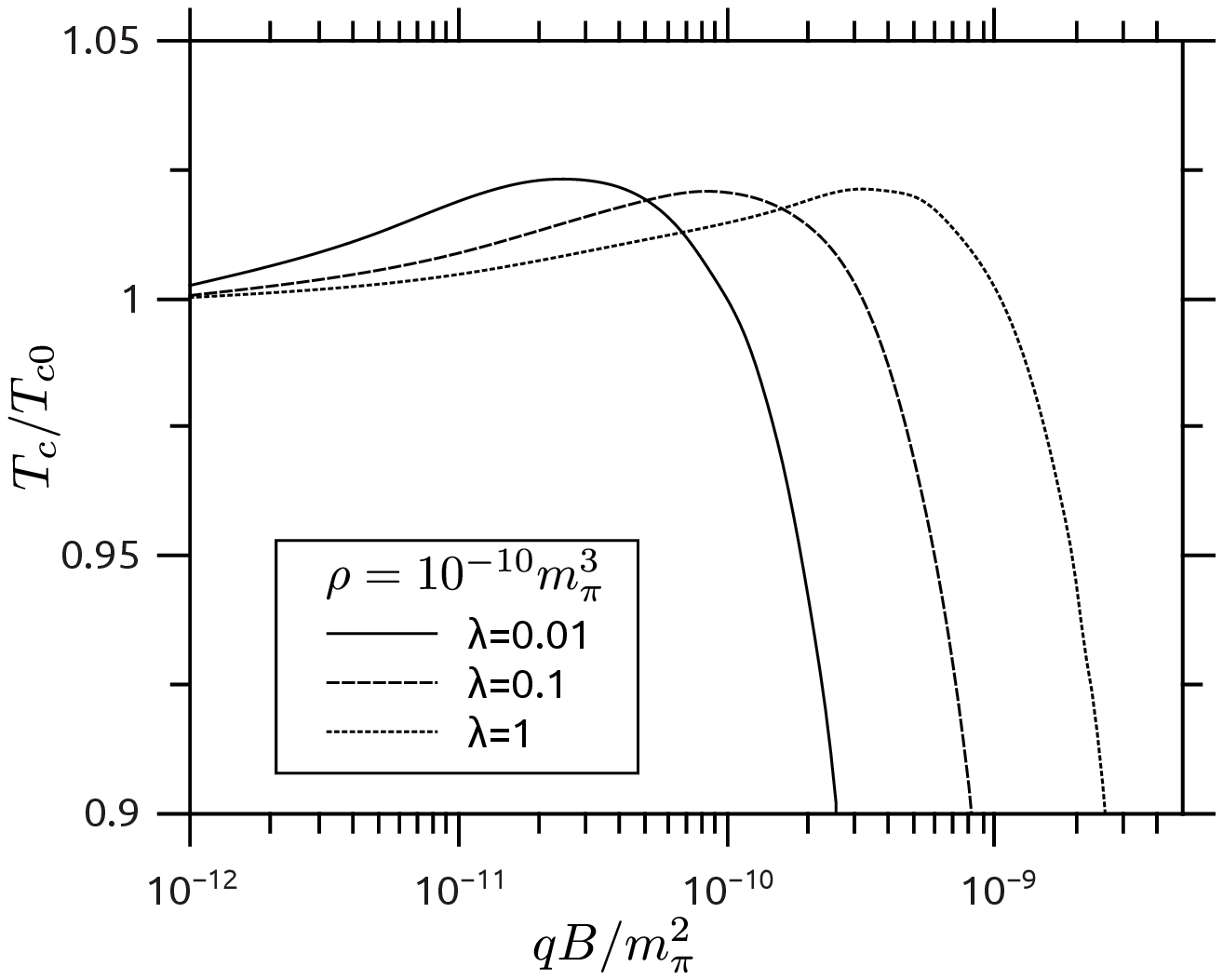}\\
   \bigskip
   \includegraphics[scale=.64]{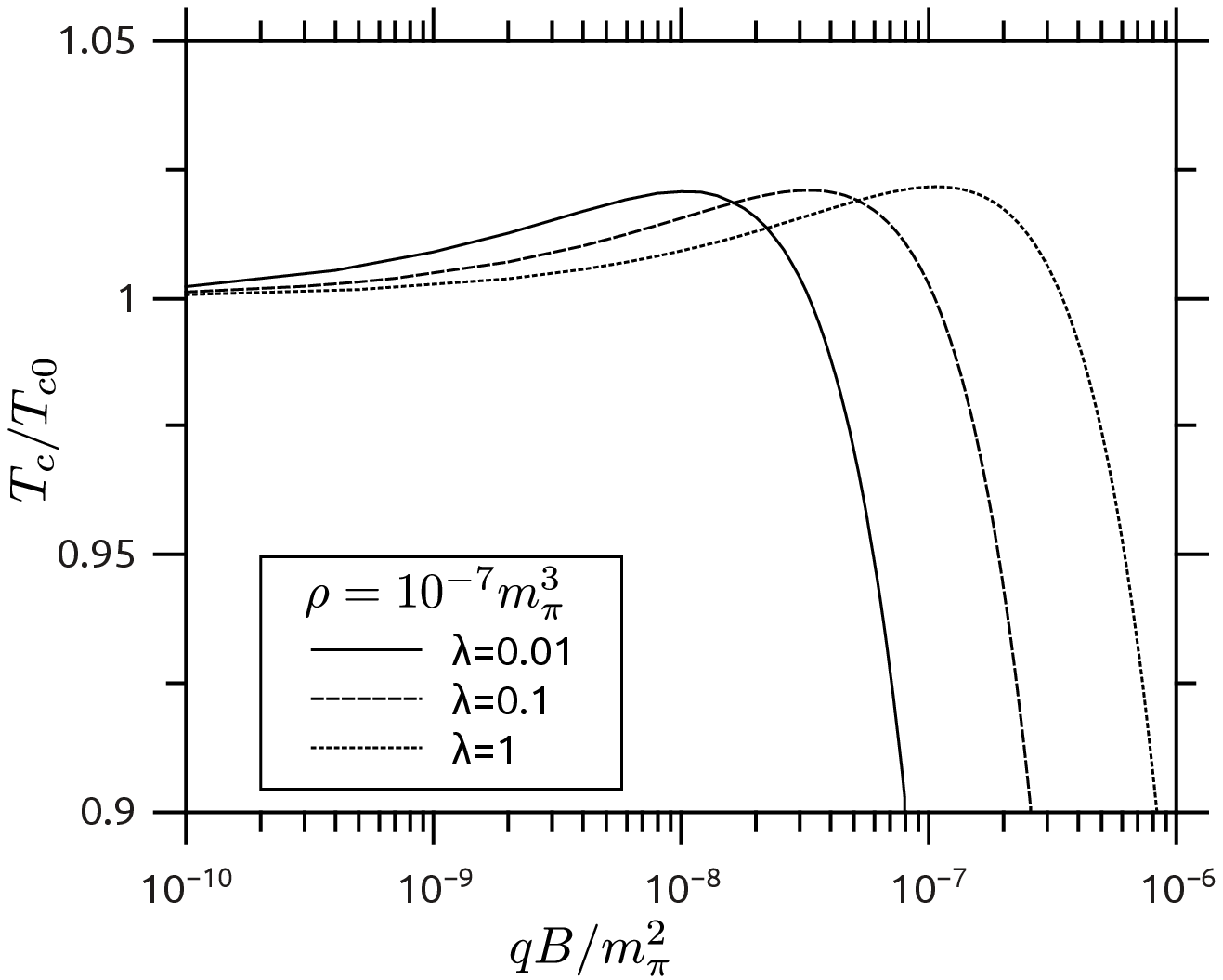}
\caption{Critical temperature $T_c$ scaled to the critical temperature $T_{c0}$
 as a function of the magnetic field strength scaled to $m_\pi^2$.
 The calculation is valid for the case when $qB\ll m_B T$.
 The upper panel corresponds to a density $\rho=10^{-10}m_\pi^3$ whereas the lower panel to a density
$\rho=10^{-7}m_\pi^3$.}
\label{fig_lowB}
\end{figure}
%%%%%%%%%%%%%%%%%%%%%%%%%%%%%%%%%%

%%%%%%%%%%%%%%%%%%%%%%%%%%%%%%%%%%
\begin{figure}
   \includegraphics[scale=.64]{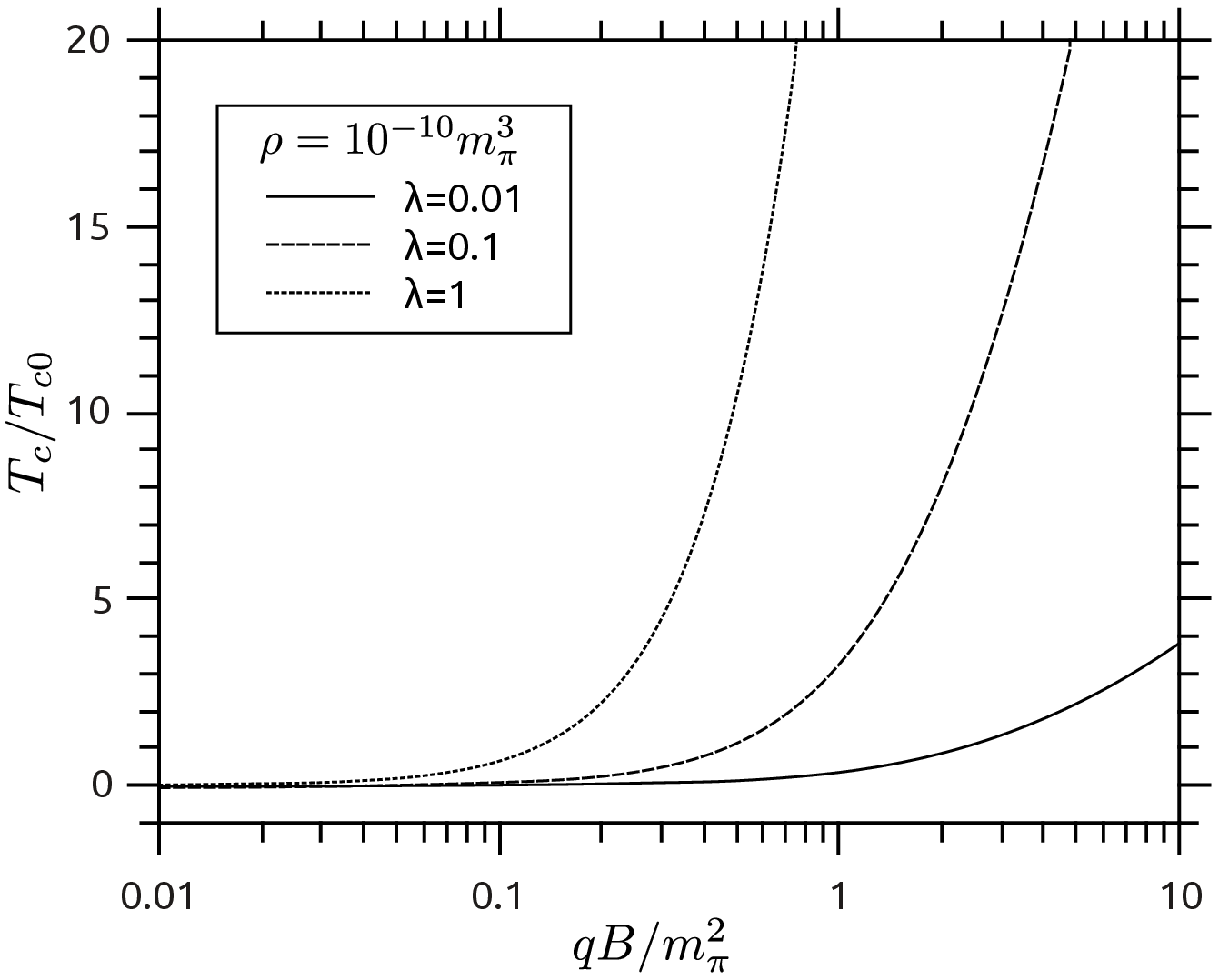}\\
   \bigskip
   \includegraphics[scale=.64]{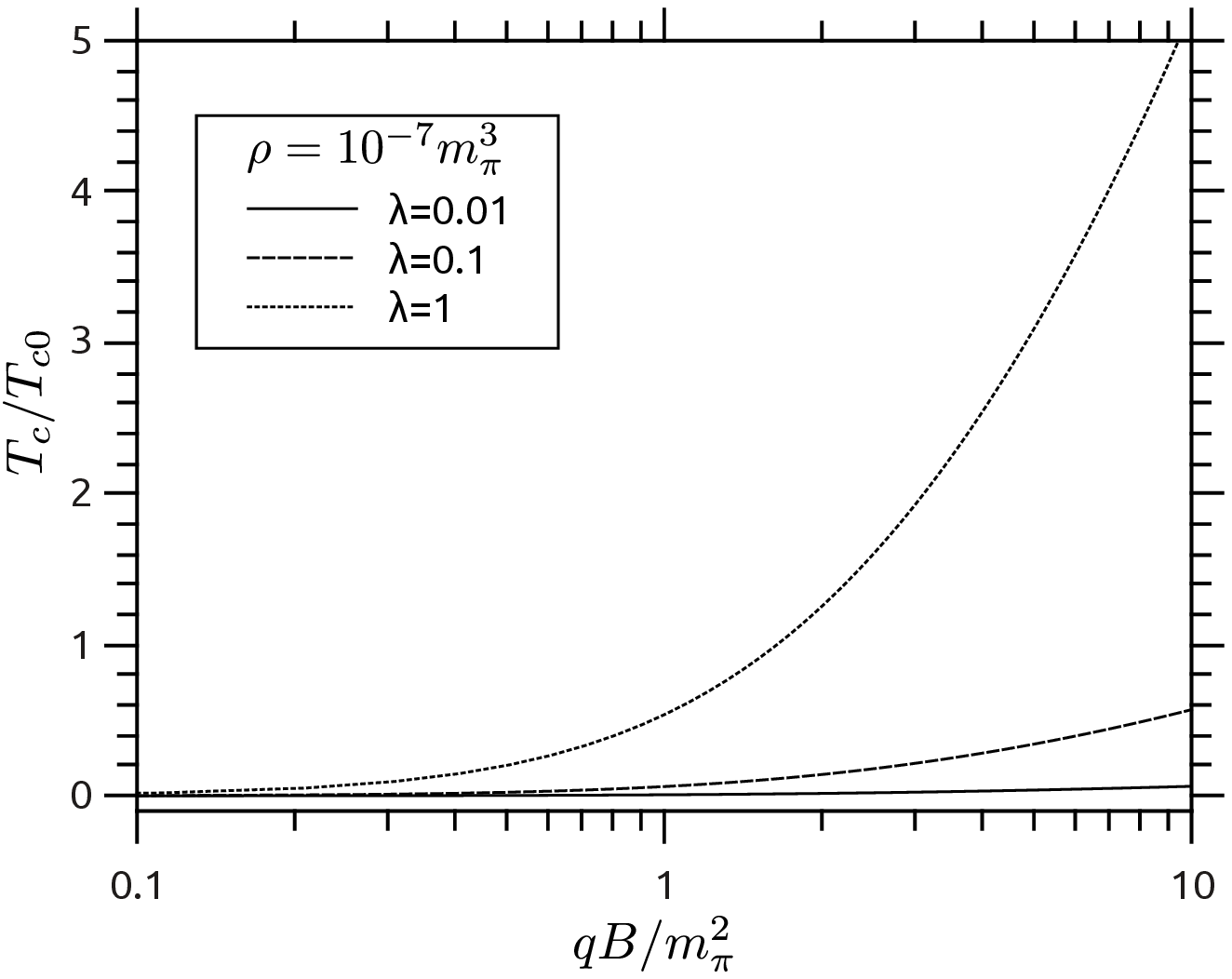}
\caption{Critical temperature $T_c$ scaled to the critical temperature $T_{c0}$ as
 a function of the magnetic field strength scaled to $m_\pi^2$.
 The calculation is valid for the case when $qB\gg m_B T$.
 The upper panel corresponds to a density $\rho=10^{-10}m_\pi^3$ whereas the lower panel to a density
$\rho=10^{-7}m_\pi^3$.}
\label{fig_highB}
\end{figure}
%%%%%%%%%%%%%%%%%%%%%%%%%%%%%%%%%%

Figures~\ref{fig_lowB} and~\ref {fig_highB} show the critical temperature
$T_c$ scaled to the critical temperature at zero magnetic field $T_{c0}$ as
a function of the magnetic field strength scaled to $m_\pi^2$. In both
cases we consider two different values of the charge density,
$\rho=10^{-10}m_\pi^3$ and $\rho=10^{-7}m_\pi^3$, which correspond roughly
to temperatures $T_0\sim 100$ eV and $T_0\sim 10$ keV, respectively.

Figure~\ref{fig_lowB} corresponds to the case $qB\ll m_B T$. In practice we
consider that the parameter $\gamma$ defined in Eq.~(\ref{scaled}) is such
that $\gamma<10^{-3}$. Notice the smooth rise of the critical temperature
as the magnetic field increases. This signals that when the magnetic field
is slowly turned on it {\it catalyzes} the formation of the condensate;
that is, the critical temperature is larger than for the $B=0$ case.
Nevertheless this behavior disappears for higher values of the magnetic
field and the formation of the condensate appears at lower temperatures. We
also mention that catalysis in this case can occur for a larger range of
magnetic field strengths when the charge density is larger, as well as for
larger values of the coupling constant $\lambda$.
Figure~\ref{fig_highB} corresponds to the case $qB\gg m_B T$. We explicitly
consider $\gamma>10^3$. 
For these larger values of $B$ the critical temperature grows. This means
that at some intermediate values of $qB$ between the large and small
regions here considered, the magnetic field catalyzes again the formation
of the condensate. The catalysis is enhanced as the coupling constant
$\lambda$ increases, and contrary to the low magnetic field case, it is
also enhanced for lower densities.

\section{Summary and conclusions}\label{VII}

In this work we have studied the condensation phenomenon for a
system of charged bosons in the presence of an external magnetic
field. Contrary to what is commonly believed, we have shown that
condensation happens at a definite critical temperature. The missing
ingredient, overlooked in previous analysis and accounted for in
this work, is the treatment of the plasma screening effects by means
of resummation. We have explicitly computed the critical
temperature, for typical densities found in compact astrophysical
objects, for small and large values of the magnetic fields. We have
shown that the magnetic field catalyzes the onset of the
condensation at very small and large values of the magnetic field,
agreeing in these regions with the case of zero chemical
potential \cite{duarte}.
 For intermediate values of the magnetic field,
  the critical
temperature for condensation turns out to be lower than in the $B=0$
case.

Recall that although the term {\it magnetic catalysis} usually refers to an enhancement of dynamical symmetry
breaking by an external magnetic field, the phenomenon seems to be
universal, appearing in different physical scenarios~\cite{Shovkovy} such
 as the one treated in this work, namely, a condensate of scalar particles, as opposed to the
more usually studied case of a fermion condensate.

We should also mention that the problem has been studied by lattice
methods as well.
The first lattice simulation for deconfinement and chiral
symmetry restoration for two-flavor QCD, in the presence of a
magnetic background, was done in Refs. \cite{D'Elia:2010nq,D'Elia:2011zu}.
The result was that the transition temperature significantly
decreased with increasing magnetic field. A similar conclusion was
presented in \cite{Ilgenfritz}. In that work it was found that in the
chirally broken phase, the chiral condensate increased monotonically
with a growing magnetic field strength. In fact, in the chiral limit
this behavior started linearly. In the same limit and in the chirally
restored phase, the condensate vanished independent of the strength
of the magnetic field. On the other hand, in
Refs.~\cite{balietal,Bali:2012zg} the effect of an external magnetic
field on the finite temperature transition of QCD was considered.
Thermodynamic observables including the chiral condensate and the
susceptibility were measured. The result was that the transition
temperature significantly decreased with increasing magnetic field.
Such discrepancies can be originated by the fact that in Refs.
\cite{D'Elia:2010nq,D'Elia:2011zu,Ilgenfritz} the pion had a large mass.
In
the case of Refs. \cite{balietal,Bali:2012zg} light pions as well as
an improved lattice were used.
 These seemingly contrasting results
call for a closer look at the phenomenon, in particular for the case
of intermediate values of the magnetic field strength. This is work
that we are currently pursuing and will be reported elsewhere.

\bigskip
\noindent

% \section*{Acknowledgments}
\begin{acknowledgements}

A.A. thanks the faculty and staff of the physics department at PUC for their
kind hospitality during a work visit in January 2012. Support for this work
has
been received in part from DGAPA-UNAM under Grant No. PAPIIT-IN103811 and
CONACyT-M\'exico under Grant No. 128534. M.L., J.C.R and C.V.
acknowledge
support from FONDECYT under Grant No. 1095217. M.L and J.C.R. acknowledge
support
from FONDECYT under Grant No. 1120770. M.L. acknowledges support from
Proyecto Anillos ACT 119 (Chile).
\end{acknowledgements}

\appendix*
 \section{}

The grand potential for a scalar field in the presence of a uniform magnetic
field evaluated at  $\phi_c=0$, is given by
\begin{eqnarray}
   \Omega(0) &=& \frac{1}{2}\text{tr} \ln ( {\cal D}_++m^2)
    +\frac{1}{2}\text{tr} \ln({\cal D}_-+m^2), \nonumber \\
   &=& \int dm^2 \; T\sum_{n=-\infty}^{\infty} \int  \frac{d^dp}{(2\pi)^d}
D(\bm{k},\omega_n),
\label{A1}
\end{eqnarray}
where the operators ${\cal D}_\pm$ are defined in Eq. (\ref{Dpm}) and the
propagator $D$ expressed in the Schwinger proper time formalism is defined in Eq. (\ref{schwprop}).
The validity of the second line in Eq.~(\ref{A1}) is more easily seen from the representation of the
grand potential in terms of Landau levels, Eq. (\ref{afterptint}).
Thus, to obtain the grand potential, we need to compute the term proportional to the one-loop
self-energy and then integrate over $m^2$, neglecting unphysical ultraviolet divergent
constants.

The sum over Matsubara frequencies $\omega_n=2\pi n T$ gives rise to the Jacobi
theta function
\[
\theta_3(z,x)=\sum_{n=-\infty}^{\infty} e^{-\pi x n^2+2\pi z n},
\]
which obeys the inversion property~\cite{temme}
\[
\theta_3(z,x)=\frac{e^{\pi z^2/x}}{\sqrt{x}} \theta_3\left(
\frac{z}{ix},\frac{1}{x} \right).
\]
Identifying $z=2 i T \mu s$ and $x=4\pi T^2 s$ allows us to rewrite the full
propagator as
\begin{eqnarray}
   T\sum_{n=-\infty}^{\infty}  D &=& \frac{1}{\sqrt{\pi}} \int_0^{\infty}
\frac{ds}{\sqrt{s}} \frac{e^{-s[m^2+p_z^2+
   p_{\perp}^2\frac{\tanh(eBs)}{eBs}]}}{\cosh(eBs)} \nonumber \\
   && \times \left[ \frac{1}{2}+\sum_{n=1}^{\infty} e^{-\frac{\beta^2
n^2}{4s}} \cosh (\beta\mu n)
\right].
   \label{Tsum_nD}
\end{eqnarray}
The ultraviolet divergent term is the first term inside the square bracket.
The one-loop self-energy, including the
counterterm $\delta m^2$, is given by
\begin{eqnarray}
   \Pi &=& \delta m^2
   +\lambda T \sum_{n=-\infty}^{\infty} \int \frac{d^3p}{(2\pi)^3}
D \nonumber \\
   &=&  \Pi_B+\Pi_{T,B},
\end{eqnarray}
where we separate the thermal contribution $\Pi_{T,B}$ from the
purely magnetic field contribution $\Pi_B$. This last contribution
contains the ultraviolet divergence, which is canceled with the counterterm
$\delta m^2$
using dimensional regularization~\cite{duarte}.
Introducing the scale factor $\Lambda$, the purely magnetic contribution is given by
\begin{eqnarray}
   \Pi_B &=& \delta m^2+\frac{\lambda}{\sqrt{4\pi}}
   \int \frac{d^dp}{(2\pi)^d} \Lambda^{3-d}\int_0^{\infty} \frac{ds}{\sqrt{s}}
   \nonumber \\
   && \times
\frac{e^{-s[m^2+p_z^2+p_{\perp}^2\frac{\tanh(qBs)}{qBs}]}}{\cosh(qBs)}.
\end{eqnarray}
Integrating the $\bm{p}_\perp$ and $p_z$ momentum components and expanding in
powers of $e^{-qBs}$, the purely magnetic contribution to the self-energy is
expressed in terms of a sum over the Landau levels
\begin{eqnarray}
  \Pi_B
   &=&
   \delta m^2+\frac{2\lambda qB \Lambda^{d-3}}{(4\pi)^{(d+1)/2}}
   \sum_{l=0}^{\infty} \int_0^\infty ds\frac{e^{-s(m^2+qB(2l+1))}}{s^{(d-1)/2}}
   \nonumber \\
   &=& \delta m^2+\frac{2\lambda qB}{(4\pi)^2}
   \left( \frac{4\pi\Lambda^2}{2qB} \right)^{(3-d)/2} \nonumber \\
   && \times
   \zeta \left( \frac{3-d}{2},\frac{1}{2}+\frac{m^2}{2qB} \right)
\Gamma\left(\frac{3-d}{2}\right),
\\&&\nonumber
\label{pib}
\end{eqnarray}
where $\zeta$ is the Hurwitz zeta function
\begin{equation}
\zeta(s,u)=\sum_{l=0}^{\infty}(l+u)^{-s},
\end{equation}
Expanding in the number of dimensions $3-d$ we get
\begin{eqnarray}
   \Pi_B &=& \delta m^2-\frac{\lambda m^2}{(4\pi)^2}
   \left[ \frac{2}{3-d}-\gamma +\ln \frac{4\pi\Lambda^2}{2qB}
   \right.
   \nonumber \\
   && \left.
   -\frac{2eB}{m^2}\zeta'\left(0,\frac{1}{2}+\frac{m^2}{2eB}\right)
   \right],
   \label{PiB_dim_reg}
\end{eqnarray}
where $\zeta'(s,u)\equiv\partial_s\zeta(s,u)$.
The divergent term is removed using $\overline{\text{MS}}$ scheme.
In order to have a vanishing contribution to the self-energy in the absence of
thermal and magnetic effect, one can choose for the scale
factor the convenient value $\Lambda=m e^{-1/2}$.
Using the relation $\zeta'(0,u)=\ln(\Gamma(u)/\sqrt{2\pi})$ we finally arrive
at the expression for $\Pi_B$ in Eq. (\ref{OmegaPiexpl}).

The expression for $\Omega_B$ in Eq. (\ref{OmegaPiexpl}) is obtained by
integrating $\Pi_B/\lambda$ with respect to $m^2$ in Eq. (\ref{PiB_dim_reg}),
and using the same value for the scale factor to remove the divergent terms in the
$\overline{\text{MS}}$ scheme.

The finite temperature part is given by the second term inside the square
bracket of Eq.~(\ref{Tsum_nD}). After integration over $\bm{p}_\perp$ and
$p_z$, the function $\Pi_{T,B}$ can be written as in Eq.~(\ref{Pi_TB}).
Consequently, by integrating $\Pi_{T,B}/\lambda$ with respect to $m^2$ we
get the function $\Omega_{T,B}$ in Eq.~(\ref{Pi_TB}).

In the limit $T \ll m_B$, one can use the steepest descent
approximation~\cite{arfken}.
By scaling the variable $s\to s'/m_B T$ in Eq. (\ref{Pi_TB}), both
integrals
can be expressed in terms of
\begin{equation}
 I=\int ds' e^{\beta m_B f(s')}g(s')
 \approx\frac{\sqrt{2\pi}g(s_0)e^{\beta
m_B f(s_0)}} {\left|{\beta m_B f''(s_0)}\right|^{1/2}}
\end{equation}
with $f(s) = -(s+n^2/4s)$ and with the saddle point $s_0=n/2$.
With this approximation we arrive at Eqs.~(\ref{Pi_TB1}) and (\ref{rho1}).

\end{document}